\documentclass[twocolumn,superscriptaddress,aps,prb]{revtex4-1}

\usepackage{bm}
\usepackage{appendix}
\usepackage{graphicx}
\usepackage{amsmath}
\usepackage{amssymb}%
\usepackage{color}
\usepackage{lipsum}
\usepackage{multirow}
\usepackage[colorlinks=true,citecolor=blue,urlcolor=blue]{hyperref}
\usepackage{mathrsfs}
\usepackage{braket}
\usepackage{hhline}

\begin{document}
	
	\title{Fate of symmetry protected coherence in open quantum system}
	
	\author{Tian-Shu Deng}
	\affiliation{Institute for Advanced Study, Tsinghua University, Beijing,100084, China}
	\author{Lei Pan}
	\email{panlei@mail.tsinghua.edu.cn}
	\affiliation{Institute for Advanced Study, Tsinghua University, Beijing,100084, China}
	\begin{abstract}
		We investigate the fate of coherence in the dynamical evolution of a symmetry protected quantum system. Under the formalism of system-plus-bath for open quantum system, the anti-unitary symmetry exhibits significant difference from the unitary one in protecting initial coherence. Specifically, taking advantage of Lindblad master equation, we find that a pure state in the symmetry protected degenerate subspace will decohere even though both the system Hamiltonian and system-environment interaction respect the same anti-unitary symmetry. In contrast, the coherence will persist when the protecting symmetry is unitary. We provide an elaborate classification table to illustrate what kinds of symmetry combinations are able to preserve the coherence of initial state, which is confirmed by several concrete models in spin-$3/2$ system. 
		Our results could help to explore the possible experimental realization of stable time-reversal symmetric topological states.

	\end{abstract}
	
	\maketitle

	
	\section{Introduction}
	Symmetry is one of the greatest unifying themes in modern physics and plays a fundamental role in classifying quantum phase of matter.
	In condensed matter physics, the existence of periodical table in topological free fermions is a great example in which the anomalous quantum Hall effect, topological insulators and topological superconductors, and many other interesting topological phenomena are unified into a single framework of classification theory. \cite{TopoClass1, TopoClass2, TopoClass3}
	Despite high degree of universality on the classification theory, the previous researches about symmetry analysis and topology in condensed matters mainly focused on the isolated systems\cite{SymClass1, SymClass2, Topo1,Topo2,Topo3,Topo4} or those systems coupled to non-Hermitian potentials\cite{NHTopo1,NHTopo2,NHTopo3,NHTopo4,NHTopo5,NHTopo6,NHTopo7,NHTopo8}. 
	
	However, realistic quantum systems inevitably couple to external degrees of freedom, which, in general, should be described by open quantum systems.
	Recently, efforts have been made to study the classification problem on the steady states in Lindblad equations\cite{Cooper2,Altland}.
	The key point for experimentally searching stable symmetry-protected topological states is maintaining the coherence of quantum state in presence of surroundings. 
	Thus, a natural question is whether these symmetry protected topological states can survive from the decoherence induced by environment.
	In particular, one of the central issues for open quantum system is the decoherence dynamics concerning how the quantum coherence evolves and vanishes which is particularly important for the quantum information and quantum computation\cite{Nielsen}. A crucial question is how to 
	avoid decoherence which is unavoidable roadblock to quantum information processing. 
	
	Therefore, diagnosing the stability of coherence of symmetry-protected quantum state has fundamental importance both in theory and practice.
	Symmetry analysis plays an important role in this issue.
	According to Wigner's theorem\cite{Wigner}, a symmetry transformation is either unitary or anti-unitary. Recent theoretical research revealed that the coherence of states underlying many symmetry protected features, could be fragile when the symmetry is anti-unitary, even though both the environment and system-environment interaction respect the same symmetry as the Hamiltonian of system\cite{Cooper}. But these features are always robust as long as the protected symmetry is unitary. 
	Strikingly, when applying this to topological systems protected by time-reversal symmetry, one concludes that the topological phases could be unstable to the perturbation of environment\cite{Cooper}. Understanding these nonequilibrium dynamics is of fundamental interest and has potential application in quantum information.
	
	In this work, we investigate the fate of coherence in degenerate subspace protected by the unitary or anti-unitary symmetry. When the system is weakly coupled to the environment, the dissipative dynamics under the Born-Markovian approximation in this open system is governed by Lindblad master equation. Indeed, the  anomalous decoherence or degeneracy breaking in anti-unitary symmetry has already arisen from the non-Hermitian linear response theory\cite{NHLRT}. At this level, dynamical evolution of density matrix in degenerate subspace is able to distinguish the difference between anti-unitary symmetry and unitary symmetry in maintaining coherence\cite{NHLRT2}. Besides, the decoherence is also related to the number of coupling channels between system and environment when the all coupling operators are Hermitian ones\cite{Cooper}. We will give a classification about the maintenance of coherence with different symmetry combinations respected by system and system-environment interaction. Furthermore, we take several representative examples in spin-$3/2$ systems to illustrate our classification. We also provide a general proof about the origin of decoherence by taking advantage of non-Hermitian Linear response theory.
	
	
	The rest of this paper is organized as follows. Sec. \ref{master} introduces the general formalism. In Sec.~\ref{Coh}, we carry out research into the roles of symmetry in dynamical evolution, and give the classification about coherence of symmetry protected states. And then in Sec.~\ref{num}, taking advantage of Lindblad master equation, we examplify the classification table with spin-$3/2$ models by calculating the density matrix and von Neumann entropy in degenerate subspace. Sec.~\ref{Summary} provides a brief summary and outlook. The analytic proof is provided in appendixes. 
	
	\section{General formalism}
	\label{master}
	In this section, we derive general formalism from microscopical Hamiltonian to study the dynamical evolution in the system protected by unitary or anti-unitary symmetry. In order to investigate the robustness of coherence in degenerate space, we consider a quantum system coupled to a Markov bath whose Hamiltonian reads
	\begin{eqnarray}
	\hat{H}_{T}&=&\hat{H}_{S}+\hat{H}_{B}+\hat{H}_{SB},
	\end{eqnarray}
	where $\hat{H}_{S}$, $\hat{H}_{B}$ are Hamiltonians belonging to the system and bath respectively, and $\hat{H}_{SB}$ denotes the interaction between them. The coupling part $\hat{H}_{SB}$ can be decomposed as 
	\begin{eqnarray}
	\hat{H}_{SB}&=&\sum_{j=1}^{M}\hat{\mathcal{A}}_{j}^{\dagger}\otimes \hat{\mathcal{B}}_{j}+\hat{\mathcal{A}}_{j}\otimes\hat{\mathcal{B}}_{j}^{\dagger},
	\end{eqnarray}
	where $\hat{\mathcal{A}}_{j}$, $\hat{\mathcal{B}}_{j}$ are operators belonging to system and bath respectively, and $M$ denotes the number of coupling channels.
	
	Let us focus on the situation that $\hat{H}_S$ has related symmetry in consideration, and then a natural question is whether the symmetry protected feature such as degeneracy is maintained or not. 
	It is generally expected that the symmetry protected feature would be destroyed if the system-bath coupling $\hat{H}_{SB}$ breaks the related symmetry and it survives as long as $\hat{H}_{SB}$ respects the same symmetry. This intuitive principle seems to be always true. 
	However, recently, in their seminal work\cite{Cooper}, McGinley and Cooper found that it can fail when the respecting symmetry is anti-unitary even each part ($\hat{\mathcal{A}}_{j}$, $\hat{\mathcal{B}}_{j}$ and $\hat{H}_{B}$) obeys the same symmetry. This unexpected discovery is deeply rooted in the Schur's Lemma for anti-unitary group\cite{anti-Schur,IrePre} and immediately indicates the fragility of time-reversal symmetry protected topological edge states\cite{NHLRT2}.
	Here we will investigate the fate of coherence in degenerate subspace of the system. For concreteness, we consider the following total Hamiltonian
	\begin{align}
	\hat{H}_{T}=\hat{H}_{S}+\sum_{j=1}^{M}\sum_{\alpha}g_{j,\alpha}\hat{\mathcal{O}}_{j}^{\dagger}\hat{b}_{\alpha}+g_{j,\alpha}^*\hat{\mathcal{O}}_{j} \hat{b}_{\alpha}^{\dagger}+\sum_{\alpha}\omega_\alpha \hat{b}_{\alpha}^{\dagger}\hat{b}_{\alpha}, 
	\end{align}
	where the bath is considered as the reservoir of harmonic oscillators in thermal equilibrium and $\hat{b}_{\alpha}$ ($\hat{b}_{\alpha}^{\dagger}$) is annihilation (creation) operator of the bath for $\alpha$ mode with bosonic commutation relation $\left[\hat{b}_{\alpha}, \hat{b}_{\beta}^{\dagger}\right]=\delta_{\alpha\beta}$ and the bath is considered as the reservoir of harmonic oscillators $\hat{H}_B=\sum_{\alpha}\omega_\alpha \hat{b}_{\alpha}^{\dagger}\hat{b}_{\alpha}$. 
	Here we focus on the scenario in which the system Hamiltonian $\hat{H}_S$ possesses the symmetry in question exhibiting degeneracy and the coupling $\hat{H}_{SB}$ and bath $\hat{H}_{B}$ respect the same symmetry. That is to say the symmetry considered here is represented by a group $\mathbf{\mathcal{G}}$ and $\hat{\mathcal{O}}$, $\hat{b}_{\alpha}$ are all invariant under the symmetry operations in $\mathbf{\mathcal{G}}$.
	
	Under this circumstance, a significant practical question is that whether symmetry protected properties such as degeneracy in system are fundamentally stable against to perturbations of the environment.
	To explore this, we consider the one-channel case $M=1$ for simplicity. 
	Following the standard procedure to integrate out the bath under Markovian approximation (low temperature bath and constant noise spectrum) and Born approximation (up to $g^2$ order), one can derive the Lindblad master equation as 
	\begin{align}
	\frac{d \hat{\rho}(t)}{d t}&\equiv\mathscr{L}\hat{\rho}(t)\nonumber \\
	&=-i\left[\hat{H}_{S}, \hat{\rho}(t)\right]-\gamma\left\{{\hat{\rho}}(t), \hat{\mathcal{O}}^{\dagger} \hat{\mathcal{O}}\right\}+2 \gamma\hat{\mathcal{O}}{\hat{\rho}}(t) \hat{\mathcal{O}}^{\dagger}, \label{Lindblad}
	\end{align}
	where $\mathscr{L}$ is Liouvillian superoperator and  $\{\hat{A},\hat{B}\}\equiv\hat{A}\hat{B}+\hat{B}\hat{A}$ denotes the anti-commutator. This equation dominates the dynamics of density matrix.
	In this way, the symmetry protected properties in degenerate space are connected with the dynamics governed by Lindblad master equation. 
	Here we focus on the von Neumann entropy in degenerate subspace to characterize decoherent process and the corresponding response of entropy is given by 
	
	\begin{eqnarray}
	\delta S(t)&=&S(t)-S_0(t)
	\end{eqnarray}
	where $S_0(t)$ is the unperturbed entropy with coherent evolution determined by $\hat{H}_S$. 
	Specifically, we will study von Neumann entropy $S_{\rm v}(t)=-{\rm Tr} [\hat{\rho}(t)\log\hat{\rho}(t)]$ which can be obtained once the master equation is solved. The decoherence dynamics in degenerate subspace and entropy growth can be derived by non-Hermitian linear response theory. Moreover, the break of degeneracy can be reflected in matrix representation of Liouvillian $\mathscr{L}$. Using the Choi-Jamio{\l}kowski isomorphism\cite{Choi,Jamiolkowski}, Eq.(\ref{Lindblad}) can be mapped to the following equation  
	\begin{align}
	\frac{d }{d t}\ket{\rho}&=\hat{\mathscr{L}}\ket{\rho},
	\label{Lindblad2}
	\end{align}
	with vectorized density matrix $\ket{\rho}$
	\begin{align}
	\ket{\rho}=\sum_{i, j} \rho_{i, j}|i\rangle \otimes|j\rangle
	\end{align} 
	where $\rho_{i,j}=\bra{i}\hat{\rho}\ket{j}$ is matrix element of $\hat{\rho}$.
	And $\hat{\mathscr{L}}$ denotes the matrix representation of Liouvillian which is written as
	\begin{eqnarray}
	\hat{\mathscr{L}}&=&-i\left(\hat{H}_S \otimes \hat{ I}-\hat{ I} \otimes \hat{H}^{\mathrm{T}}_S\right)\nonumber \\
	&+&\gamma\left[2\hat{\mathcal{O}} \otimes \hat{\mathcal{O}}^{*}- \hat{\mathcal{O}}^{\dagger} \hat{\mathcal{O}}\otimes \hat{ I}-\hat{ I}\otimes\left(\hat{\mathcal{O}}^{\dagger} \hat{\mathcal{O}}\right)^{\mathrm{T}}  \right]. \label{L_G}
	\end{eqnarray}
	The Liouvillian superoperator shares the symmetry associated with $\hat{\mathcal{O}}$ which determines the symmetry of Eq.(\ref{Lindblad}).     
	The fate of coherence in degenerate subspace depends on whether $\hat{\mathscr{L}}$ is proportional to identity matrix in degenerate subspace.
	
	
	\section{Coherence analysis for different symmetry combinations} 
	\label{Coh}
	In this section, we classify the coherent dynamics in degenerate subspace  regarding different combinations of symmetries respected by $\hat{H}_S$ and the $\hat{\mathcal{O}}$.
	Supposing the system respects unitary or anti-unitary symmetry characterized by group $\mathbf{\mathcal{G}}$. This means the matrix representation of $\hat{H}_S$ satisfies $[\hat{H}_S,\hat{U}_\mathbf{\mathcal{G}}]=0$ for any group element $\hat{U}_\mathbf{\mathcal{G}}\in \mathbf{\mathcal{G}}$. If $|\psi\rangle$ is one of eigenstates of $\hat{H}_S$, i.e.,   $\hat{H}_S|\psi\rangle=E|\psi\rangle$, then  $\hat{U}_\mathbf{\mathcal{G}}|\psi\rangle$ would also be the eigenstate of $\hat{H}_S$. Thus $\{\hat{U}_\mathbf{\mathcal{G}}|\psi\rangle\}$ spans a degenerate subspace when $\hat{U}_\mathbf{\mathcal{G}}\in \mathbf{\mathcal{G}}$ satisfying $\hat{U}_\mathbf{\mathcal{G}}|\psi\rangle\neq|\psi\rangle$ exists, which is also irreducible representation subspace of $\mathbf{\mathcal{G}}$. We call this  symmetry protected degeneracy. A notable example in half-odd integer spin  system is the Kramers' degeneracy when $\mathbf{\mathcal{G}}$ represents time reversal symmetry group.
	
	In order to describe the coherent dynamics in degenerate space, we assume that the $\{U_G|\psi\rangle\}$ contains two orthogonal basis which are denoted by $|\phi_+\rangle$ and $|\phi_-\rangle$ corresponding to two-fold degenerate subspace.
	The initial state is prepared to be a pure state $\hat{\rho}(0)=|\psi(0)\rangle\langle\psi(0)|$ with $|\psi(0)\rangle=\alpha|\phi_+\rangle+\beta|\phi_-\rangle$ and then the dynamics of density matrix will be dominated by Lindblad master equation (\ref{Lindblad}). We focus on the density matrix in the subspace $\hat{\rho}_{G}(t)=\hat{\Pi}_{G}\hat{\rho}(t)\hat{\Pi}_{G}$, where the ground state projective operator is defined by  $\hat{\Pi}_{G}=|\phi_+\rangle\langle\phi_+|+|\phi_-\rangle\langle\phi_-|$.
	
	For a general total Hamiltonian, the bath and system-bath coupling don't respect the related symmetry of $\hat{H}_S$, in which case $\hat{\rho}_G(t)$ evolves into a mixed state where the decoherence happens. Nevertheless, what we concern here is whether the complete information of initial state, or at least the coherence, can be maintained when both the operators \{$\hat{a}_\alpha$\} and \{$\hat{\mathcal{O}}_{j}$\} respect the same symmetry as the system Hamiltonian $\hat{H}_S$.
	As stated above, the fate of coherence is determine by unitarity or anti-unitarity of the symmetry. This conclusion had been pointed out in Ref.\cite{Cooper,NHLRT2} and here we provide a brief explanation. 
	For the unitary symmetry, $\hat{\Pi}_{G}\hat{\mathcal{O}}^\dagger\hat{\mathcal{O}}\hat{\Pi}_{G}\propto \hat{\Pi}_{G}$,  $\hat{\Pi}_{G}\hat{\mathcal{O}}^\dagger\hat{\Pi}_{G}\propto \hat{\Pi}_{G}$, and $\hat{\Pi}_{G}\hat{\mathcal{O}}\hat{\Pi}_{G}\propto \hat{\Pi}_{G}$ are guaranteed by Schur's lemma\cite{Schur} as long as $\hat{\mathcal{O}}^\dagger$, $\hat{\mathcal{O}}$ respect the same symmetry as $\hat{H}_S$. In this case, Lindblad superoperator $\mathscr{L}$ only acts on the subspace density matrix as $\hat{\Pi}_{G}\mathscr{L}[\hat{\rho}(t)]\hat{\Pi}_{G}\propto \rho_{G}(0)$.
	Accordingly, if we define
	\begin{equation}
	\rho_{G}(t)=\left(\begin{array}{cc}
	\rho_{++} & \rho_{+-}\\
	\rho_{-+} & \rho_{--}
	\end{array}\right)=\left(\begin{array}{cc}
	\langle\phi_{+}|\hat{\rho}|\phi_{+}\rangle & \langle\phi_{+}|\hat{\rho}|\phi_{-}\rangle\\
	\langle\phi_{-}|\hat{\rho}|\phi_{+}\rangle & \langle\phi_{-}|\hat{\rho}|\phi_{-}\rangle
	\end{array}\right),
	\end{equation}
	then all of the matrix elements synchronously decay with the same rate. That is to say, renormalized density matrix in subspace $\tilde{\rho}_G=\rho_G(t)/{\rm tr}(\rho_G(t))$ is always equal to $\rho_G(0)$ under the time-evolution of Lindblad equation which means the initial coherence is maintained.
	In addition, the matrix representation of $\hat{\mathscr{L}}$ in subspace  
	is proportional to identity, namely, 
	\begin{align}
	&\hat{\mathscr{L}}_G\propto\left(\begin{array}{cccc}
	1 & 0 & 0 & 0\\
	0 & 1 & 0 & 0\\
	0 & 0 & 1 & 0\\
	0 & 0 & 0 & 1
	\end{array}\right),
	\label{L_G1}
	\end{align}
	which acts trivially on the state of subspace $\hat{\mathscr{L}}_G\big[\rho_{+,+}(t),\rho_{+,-}(t),\rho_{-,+}(t),\rho_{-,-}(t)\big]^{\rm T}\propto\big[\rho_{+,+}(0),\rho_{+,-}(0),\rho_{-,+}(0),\rho_{-,-}(0)\big]^{\rm T}$.   
	For the case of anti-unitary symmetry, according to the Schur's Lemma\cite{anti-Schur}, if $\hat{\mathcal{O}}$ is Hermitian operator obeying this symmetry, it will be proportional to identity matrix in degenerate subspace, i.e.,  $\hat{\Pi}_{G}\hat{\mathcal{O}}^\dagger\hat{\Pi}_{G}=\hat{\Pi}_{G}\hat{\mathcal{O}}\hat{\Pi}_{G}\propto \hat{I}_{G}$ and the condition (\ref{L_G1}) is also true which means the density matrix will maintain its initial coherence. But if $\hat{\mathcal{O}}$ is non-Hermitian operator, then its matrix presentation is proportional to identity, i.e.,  $\hat{\Pi}_{G}\hat{\mathcal{O}}\hat{\Pi}_{G}\not\propto \hat{I}_{G}$. In this case, one can find immediately $\tilde{\rho}_G=\rho_G(t)/{\rm tr}(\rho_G(t))\neq\rho_G(0)$ from master equation (\ref{Lindblad}) and decoherence occurs. Meanwhile $\hat{\mathscr{L}}_G$ is no longer proportional to identity 
	\begin{align}
	&\hat{\mathscr{L}}_G\not\propto\left(\begin{array}{cccc}
	1 & 0 & 0 & 0\\
	0 & 1 & 0 & 0\\
	0 & 0 & 1 & 0\\
	0 & 0 & 0 & 1
	\end{array}\right).
	\label{L_G2}
	\end{align}
	%
	%
	The above statements for unitary or anti-unitary symmetric systems can be naturally generalized to those with various combinations of symmetries.
	We summarize different kinds of symmetry combinations and corresponding fate of coherence in Table~\ref{class}, we take time-reversal symmetric group and quaternion group as candidates of antiuntary and unitary group, respectively. Some representative examples  will be illustrated in next section. 
	\begin{widetext}

		\begin{table}
			\centering
			\caption{Classification table for the fate of initial coherence protected by there kinds of symmetries in $\hat{H}_s$ and coupled to operator $\mathcal{O}$ with different types of symmetry combinations. Here $[\hat{H}_S,\mathcal{Q}]=0$ ($[\hat{H}_S,\mathcal{T}]=0$) means the system respects ($\mathcal{Q}$-symmetry) time-reversal symmetry. $\hat{I}_G$ represents the identity matrix in ground-state subspace.}
			\label{class}

			
			\begin{tabular}{|c|c|c|c|c|} 
				
				\hline
				Symmetry of $\hat{H}_S$                                                                                                                              & Hermiticity of $\hat{\mathcal{O}}$  & Symmetry of $\hat{\mathcal{O}}$                                                       & Coherence/Decoherence & $\hat{\mathscr{L}}_G\propto\hat{I}_G\otimes\hat{I}_G$ (Yes/No)   \\ 
				\hline
				\multirow{4}{*}{$[\hat{H}_S,\mathcal{Q}]=0$}                                                                                                         & \multirow{2}{*}{Hermitian}          &  $[\hat{\mathcal{O}},\mathcal{Q}]=0$      & Coherence             & Yes                                                              \\ 
				\cline{3-5}
				&                                     & $[\hat{\mathcal{O}},\mathcal{Q}] \neq 0$                                              & Decoherence           & No                                                               \\ 
				\cline{2-5}
				& \multirow{2}{*}{Non-Hermitian}      & $[\hat{\mathcal{O}},\mathcal{Q}] = 0$                                                 & Coherence           & Yes                                                               \\ 
				\cline{3-5}
				&                                     & $[\hat{\mathcal{O}},\mathcal{Q}] \neq 0$                                              & Decoherence           & No                                                               \\ 
				\hline
				\multirow{4}{*}{$[\hat{H}_S,\mathcal{T}]=0$ }                                                                                                         & \multirow{2}{*}{Hermitian}          & $[\hat{\mathcal{O}},\mathcal{T}] = 0$                                                 & Coherence             & Yes                                                              \\ 
				\cline{3-5}
				&                                     & $[\hat{\mathcal{O}},\mathcal{T}] \neq 0 $                                             & Decoherence           & No                                                               \\ 
				\cline{2-5}
				& \multirow{2}{*}{Non-Hermitian}      & $[\hat{\mathcal{O}},\mathcal{T}] = 0$                                                 & Decoherence             & No                                                              \\ 
				\cline{3-5}
				&                                     & $[\hat{\mathcal{O}},\mathcal{T}] \neq 0$                                              & Decoherence           & No                                                               \\ 
				\hline
				\multirow{8}{*}{ $\begin{array}{c}[\hat{H}_{S},\mathcal{T}]=0\\{}[\hat{H}_{S},\mathcal{Q}]=0\end{array}$ } & \multirow{4}{*}{Hermitian}          & $[\hat{\mathcal{O}},\mathcal{T}] = 0,\quad[\hat{\mathcal{O}},\mathcal{Q}] = 0$        & Coherence             & Yes                                                              \\ 
				\cline{3-5}
				&                                     & $[\hat{\mathcal{O}},\mathcal{T}] = 0,\quad[\hat{\mathcal{O}},\mathcal{Q}] \neq 0$     & Coherence             & Yes                                                              \\ 
				\cline{3-5}
				&                                     & $[\hat{\mathcal{O}},\mathcal{T}] \neq 0,\quad[\hat{\mathcal{O}},\mathcal{Q}] = 0$     & Coherence             & Yes                                                              \\ 
				\cline{3-5}
				&                                     & $[\hat{\mathcal{O}},\mathcal{T}] \neq 0,\quad[\hat{\mathcal{O}},\mathcal{Q}] \neq 0$  & Decoherence           & No                                                               \\ 
				\cline{2-5}
				& \multirow{4}{*}{Non-Hermitian}      & $[\hat{\mathcal{O}},\mathcal{T}] = 0,\quad[\hat{\mathcal{O}},\mathcal{Q}] = 0$        & Coherence             & Yes                                                              \\ 
				\cline{3-5}
				&                                     & $[\hat{\mathcal{O}},\mathcal{T}] = 0,\quad[\hat{\mathcal{O}},\mathcal{Q}] \neq 0$     & Decoherence           & No                                                               \\ 
				\cline{3-5}
				&                                     & $[\hat{\mathcal{O}},\mathcal{T}] \neq 0,\quad[\hat{\mathcal{O}},\mathcal{Q}] = 0$     & Coherence             & Yes                                                              \\ 
				\cline{3-5}
				&                                     & $[\hat{\mathcal{O}},\mathcal{T}] \neq 0,\quad[\hat{\mathcal{O}},\mathcal{Q}]\neq 0$   & Decoherence           & No                                                               \\
				\hline
			\end{tabular}
			
		\end{table}                                    
	\end{widetext}

	
	\section{Numerical results with spin-$3/2$ models}
	\label{num}
	In this section, we take spin-$3/2$ models as a concrete example to elucidate and verify the symmetry classification shown in Table \ref{class}. This kind of high-spin systems coupled to environment was extensively studied in nitrogen-vacancy (NV) centers\cite{NVcenter1,NVcenter2}.
	The matrix representation of spin-3/2 angular momentum is chosen as 
	\begin{align}
	&S_{x}=\frac{1}{2}\left(\begin{array}{cccc}
	0 & \sqrt{3} & 0 & 0\\
	\sqrt{3} & 0 & 2 & 0\\
	0 & 2 & 0 & \sqrt{3}\\
	0 & 0 & \sqrt{3} & 0
	\end{array}\right),\nonumber\\
	&S_{y}=\frac{i}{2}\left(\begin{array}{cccc}
	0 & -\sqrt{3} & 0 & 0\\
	\sqrt{3} & 0 & -2 & 0\\
	0 & 2 & 0 & -\sqrt{3}\\
	0 & 0 & \sqrt{3} & 0
	\end{array}\right),\nonumber\\
	&S_{z}=\left(\begin{array}{cccc}
	\frac{3}{2} & 0 & 0 & 0\\
	0 & \frac{1}{2} & 0 & 0\\
	0 & 0 & -\frac{1}{2} & 0\\
	0 & 0 & 0 & -\frac{3}{2}
	\end{array}\right).
	\end{align}
	
	The unitary and anti-unitary symmetries will be discussed respectively.
	
	\subsection{Unitary symmetry}\label{Uni-sym}
\begin{figure}[!h]
		\centering
		\includegraphics[width=8.0cm]{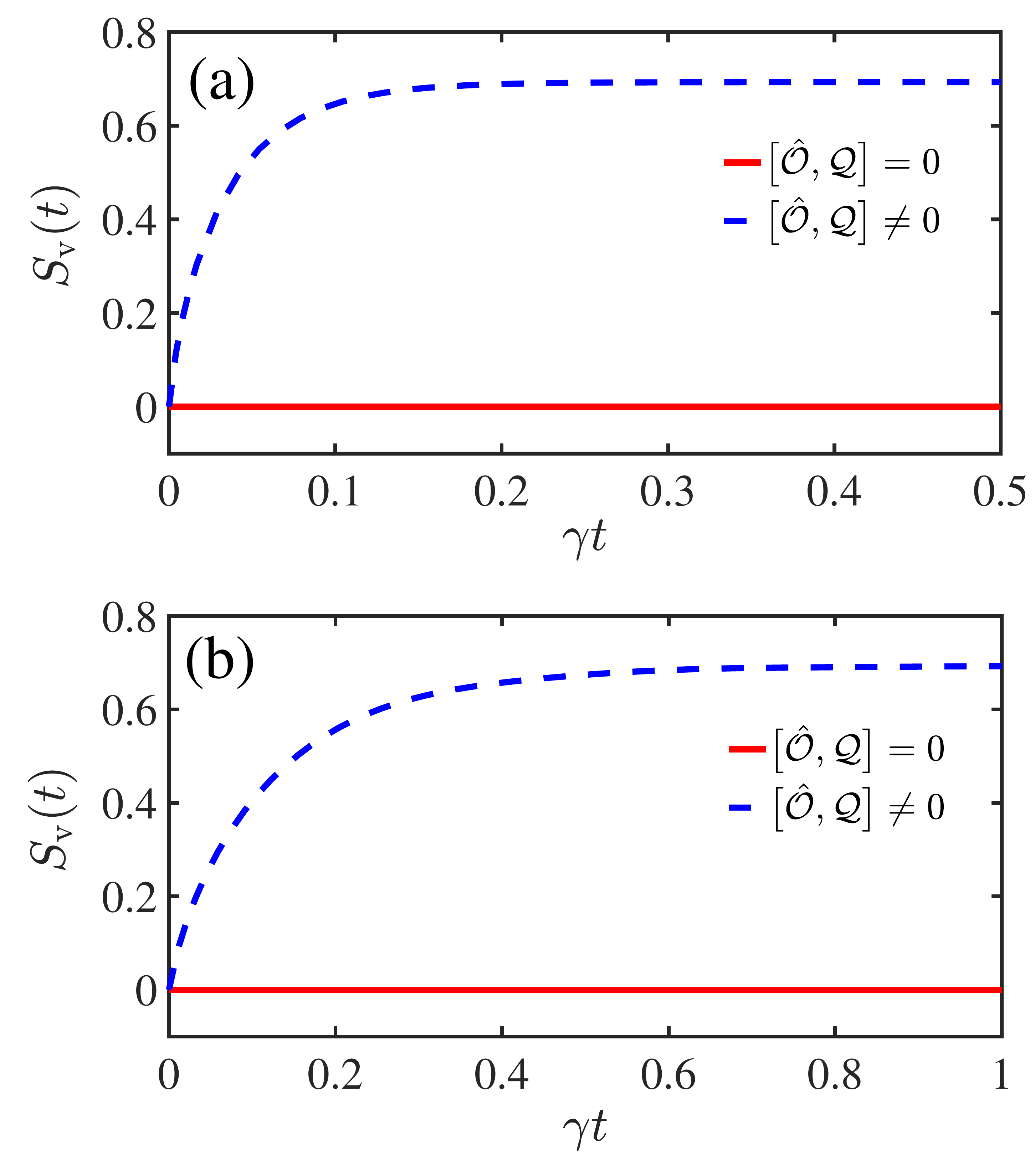}
		\caption{Time evolution of von Neumann entropy in degenerate subspace protected by unitary-symmetry ($\mathcal{Q}$-symmetry). (a) Von Neumann entropy as function of time for hermitian coupling operator: with $\mathcal{Q}$-symmetry $\hat{Q}=\hat{S}_y^2$ (red solid line) and without $\mathcal{Q}$-symmetry $\hat{Q}=\hat{S}_x\hat{S}_y+\hat{S}_y\hat{S}_x$ (blue dashed line). (b) Von Neumann entropy as function of time for non-hermitian coupling operator: with $\mathcal{Q}$-symmetry $\hat{Q}=\hat{S}_x\hat{S}_y\hat{S}_z$ (red solid line) and without $\mathcal{Q}$-symmetry $\hat{Q}=\hat{S}_x\hat{S}_y\hat{S}_z$ (blue dashed line). }
		\label{Fig1}
	\end{figure}

	We first study the system with unitary symmetry. In order illustrate the symmetry protected coherence, for simplicity, we choose the quaternion group $\mathcal{Q}$ as candidate of the unitary group which is a non-Abelian unitary group whose irreducible representation can be two dimensions, spanning a two-fold degenerate subspace. The matrix representation of $\mathcal{Q}$ is displayed explicitly in Appendix \ref{App2}. We then construct a $\mathcal{Q}$-symmetric Hamiltonian $\hat{H}_{S}=E_{g}(\hat{S}_{x}\hat{S}_{y}\hat{S}_{z}+\hat{S}_{z}\hat{S}_{y}\hat{S}_{x})$ with twofold degenerate ground states which forms the two-dimensional irreducible representation space of group $\mathcal{Q}$. For Hermitian operator $\hat{\mathcal{O}}$ coupled to system respecting $\mathcal{Q}$-symmetry such as $\hat{\mathcal{O}}=\hat{S}_{y}^2$, the coherence in ground-state subspace maintains all the time but decoherence occurs in the case of breaking this symmetry, say, $\hat{\mathcal{O}}=\hat{S}_{x}\hat{S}_{y}+\hat{S}_{y}\hat{S}_{x}$ as shown in Fig.\ref{Fig1}(a). The same situation happens to the non-Hermitian coupling operators that the symmetric and asymmetric $\hat{\mathcal{O}}$ are chosen as $\hat{\mathcal{O}}=\hat{S}_{x}\hat{S}_{y}\hat{S}_{z}$ and $\hat{\mathcal{O}}=\hat{S}_{y}\hat{S}_{z}$ respectively (see Fig.\ref{Fig1}(b)). The increasing entropy reflects the fact subspace density matrix evolves into a mixed state. The unchanged von Neumann entropy indicates that the density matrix in subspace is always a pure state, which means the coherence is maintained. This is reasonable consequence that the decoherence appears only when the system is perturbed by symmetry-broken coupling. As mentioned above, the Schur's lemma allows us to infer that $\hat{\Pi}_{G}\hat{\mathcal{O}}\hat{\Pi}_{G}\propto \hat{I}_{G}$ if $\hat{\mathcal{O}}$ respects $\mathcal{Q}$-symmetry which will lead to $\hat{\mathscr{L}}_G\propto\hat{I}_G\otimes\hat{I}_G$ according to expression (\ref{L_G}). In contrast, $\hat{\mathscr{L}}_G\not\propto\hat{I}_G\otimes\hat{I}_G$ if $\hat{\Pi}_{G}\hat{\mathcal{O}}\hat{\Pi}_{G}\not\propto \hat{I}_{G}$ since $\hat{\mathcal{O}}$ breaks $\mathcal{Q}$-symmetry which destroys the coherence and the corresponding entropy of decoherent dynamics reaches steady-state value $S_{\rm v}(\infty)=\ln 2\approx0.693$ which is nothing but the maximum entropy in doublet space. 
	\subsection{Anti-unitary symmetry}\label{Anti-Uni-sym}
\begin{figure}[h]
		\centering
		\includegraphics[width=8.5cm]{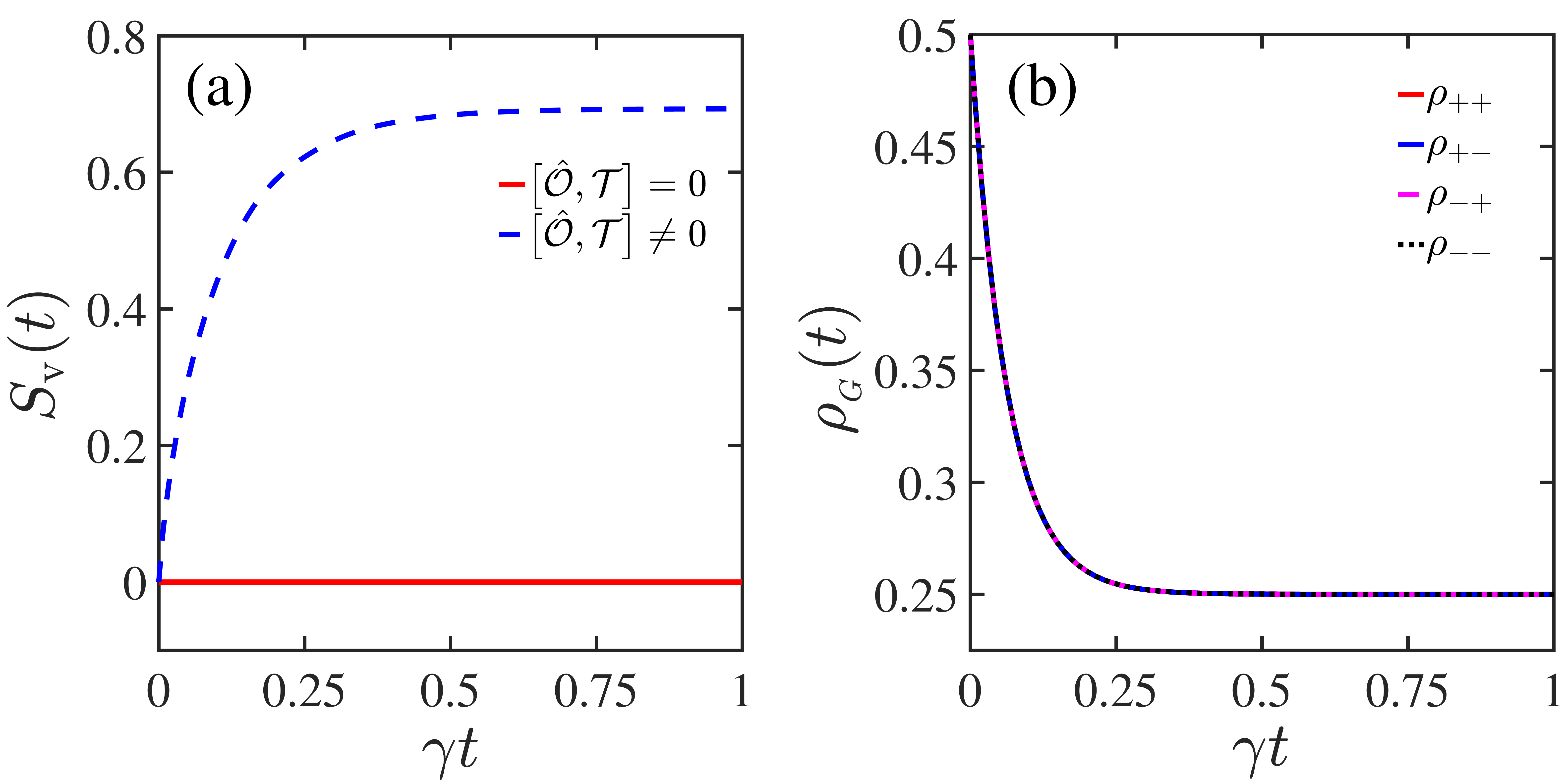}
		\includegraphics[width=8.5cm]{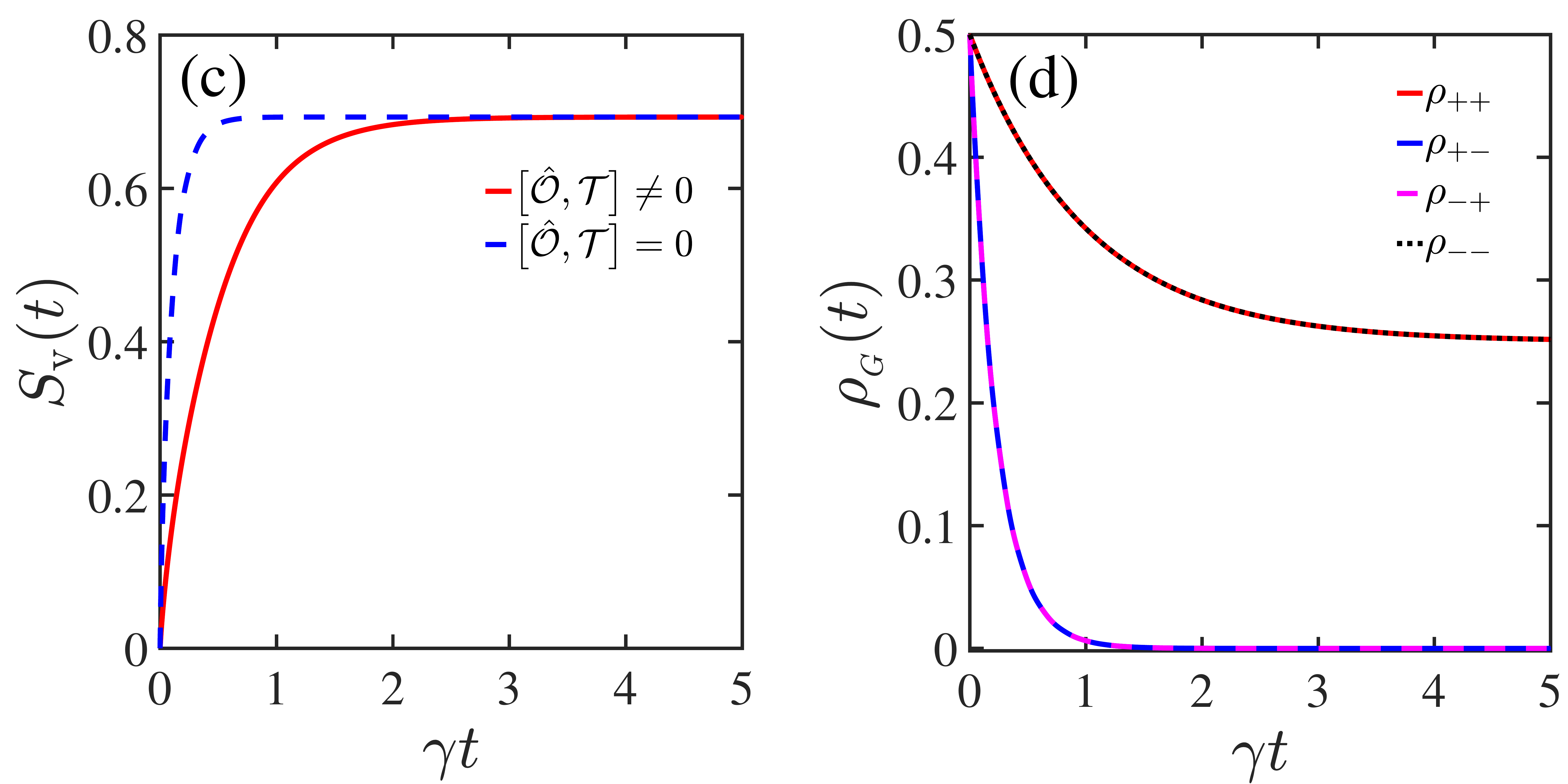}
		\caption{Time evolution of von Neumann entropy and density matrix in degenerate subspace protected by time-reversal symmetry. (a) Von Neumann entropy as function of time for hermitian coupling operator: with time-reversal symmetry $\hat{Q}=S_x^2$ (red solid line) and without time-reversal symmetry $\hat{Q}=S_z$ (blue dashed line). (b) Matrix elements of density matrix associated with red solid line in (a) evolves as time. (c) Von Neumann entropy as function of time for non-Hermitian coupling operator: without time-reversal symmetry $\hat{Q}=S_xS_yS_z$ (red solid line) and with time-reversal symmetry $\hat{Q}=iS_z$ (blue dashed line). (d) Matrix elements of density matrix associated with blue dashed line in (c) evolves as time. The initial value of density matrix is chosen as $\ket{\psi(0)}\bra{\psi(0)}$ with $\ket{\psi(0)}=\frac{1}{\sqrt{2}}\big(\ket{\phi_+}+\ket{\phi_-}\big)$.}\label{Fig2}
	\end{figure}

	From now on, we focus on the systems respecting time-reversal symmetry. Since time-reversal operation inverts the angular momentum operators, i.e., $\hat{T}\hat{S}_{x,y,z}\hat{T}^{-1}=-\hat{S}_{x,y,z}$, one can construct time-reversal invariant Hamiltonian as $\hat{H}_S=E_{g}\{\hat{S}_x,\hat{S}_z\}$. The ground-state subspace of $\hat{H}_S$ is two-fold degenerate due to Kramers' theorem and we denote two degenerate states as $|\phi_\pm\rangle$. Similar to the discussion of unitary symmetric system, here the hermitian and non-hermitian coupling $\hat{\mathcal{O}}$ will be investigated respectively. 
	In order to verify the case of $[\hat{H}_S,\mathcal{T}]=0$ in Table \ref{class}, we prepare the initial state on $|\psi(0)\rangle=\alpha|\phi_+\rangle+\beta|\phi_-\rangle$, and then calculate the time evolution of von Neumann entropy in ground-state subspace by means of Lindblad equation shown in Fig.~\ref{Fig2}. 
	
	For Hermitian case, as shown in Fig.\ref{Fig2}(a), the initial coherence will maintain (vanish) if the operator $\hat{\mathcal{O}}$ coupled by system respects (breaks) time-reversal symmetry. However, there will be a dramatic difference when this time-reversal symmetric system couples to non-Hermitian operator $\hat{\mathcal{O}}$. Fig.\ref{Fig2}(c) plots the von Neumann entropy from which one can see clearly that the coherence is always destroyed despite whether or not $\hat{\mathcal{O}}$ respect time-reversal symmetry. Hence, we demonstrate the statement that coherence could survive only if the time-reversal symmetric coupling operator is also Hermitian.
	This is consistent with Schur's Lemma for anti-unitary group that an anti-unitary symmetric and hermitian operator is proportional to identity in degenerate subspace, i.e., $\hat{\Pi}_{G}\hat{\mathcal{O}}\hat{\Pi}_{G}\propto \hat{I}_{G}$ which makes the relation $\hat{\mathscr{L}}_G\propto\hat{I}_G\otimes\hat{I}_G$ true. When the situation that any of symmetry and hermiticity is not satisfied happens, it results in $\hat{\Pi}_{G}\hat{\mathcal{O}}\hat{\Pi}_{G}\not\propto \hat{I}_{G}$ which causes the decoherence and meanwhile $\hat{\mathscr{L}}_G\not\propto\hat{I}_G\otimes\hat{I}_G$ .
	
	With regard to density matrix $\hat{\rho}_G(t)$ in ground-state space, for coherent evolution, all matrix elements decay with equal-rate during whole region of time as shown in Fig.\ref{Fig2}(b). While for decoherent process, the nondiagonal elements could decay to zero but its diagonal elements keep finite which means all coherence is lost and the system reaches maximum entanglement in groundstate subspace (see Fig.\ref{Fig2}(d)). 
	
	\subsection{Both unitary and anti-unitary symmetry}\label{Uni-Anti}
\begin{figure}[!h]
		\centering
		\includegraphics[width=8.2cm]{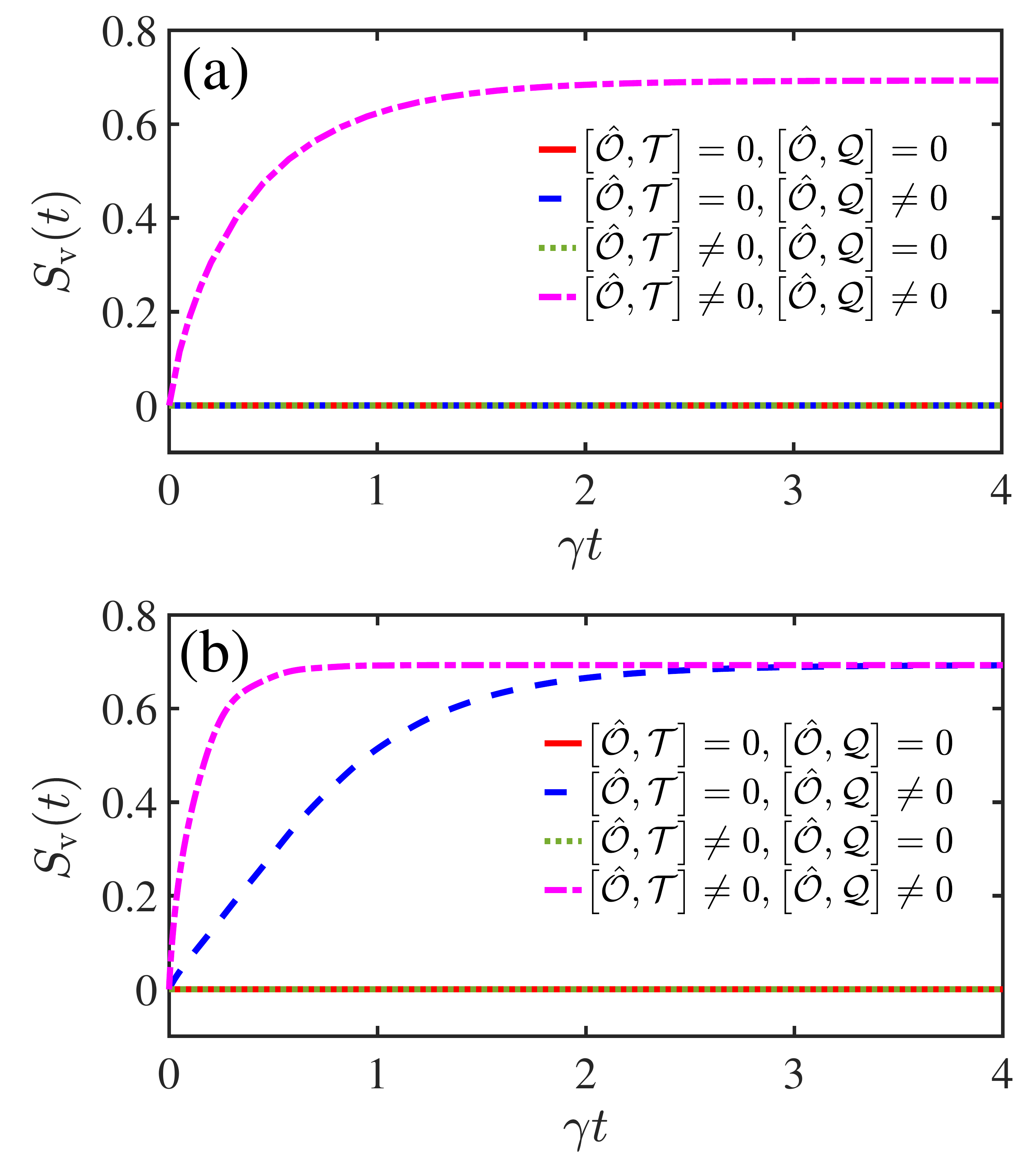}
		\caption{The von Neumann entropy as a function of time with the system Hamiltonian respecting both $\mathcal{Q}$-symmetry and time-reversal symmetry. The Hermitian and non-Hermitian coupling $\hat{\mathcal{O}}$'s are shown in (a) and (b) respectively. 
			For hermitian case, the four kinds of coupling $\hat{\mathcal{O}}$ are chosen as $\hat{S}_x^2$ (red solid line), $\hat{S}_x\hat{S}_y+\hat{S}_y\hat{S}_x$ (blue dashed line),  $\hat{S}_x\hat{S}_y\hat{S}_z+\hat{S}_z\hat{S}_y\hat{S}_x$ (green dotted line), and $\hat{S}_x$ (dot-dashed line). For non-Hermitian case, the corresponding choice is $i(\hat{S}_x\hat{S}_y\hat{S}_z-\hat{S}_z\hat{S}_y\hat{S}_x)$, $\hat{S}_x\hat{S}_y$, $\hat{S}_x\hat{S}_y\hat{S}_z$, and $\hat{S}_x^2\hat{S}_z$. The initial value of density matrix is chosen as $\ket{\psi(0)}\bra{\psi(0)}$ with $\ket{\psi(0)}=\frac{1}{\sqrt{2}}\big(\ket{\phi_+}+\ket{\phi_-}\big)$.}
		\label{Fig3}
	\end{figure}

	There is another case in which the Hamiltonian respects both time-reversal symmetry and $\mathcal{Q}$-symmetry, such as $\hat{H}_S=E_g\hat{S}_z^2$ whose degeneracy of groundstates is protected by both of symmetries. This highly symmetric system allows for more types of symmetry combinations which contain four circumstances for both Hermitian and non-Hermitian couplings as shown in the third row of Table \ref{class}. The corresponding entropy dynamics associated with different symmetries is shown in Fig.~(\ref{Fig3}).
	For hermitian case, the initial coherence in ground-state subspace is always maintained whichever symmetry the coupling operator $\hat{\mathcal{O}}$ has. This is consistent with the conclusions in subsection \ref{Uni-sym} and \ref{Anti-Uni-sym} that $\hat{\Pi}_{G}\hat{\mathcal{O}}\hat{\Pi}_{G}\propto \hat{I}_{G}$ regardless of unitarity or antiunitarity which signifies the decoherence occurs only when the coupling operator breaks both symmetries, as shown in Fig.\ref{Fig3}.(a). For non-Hermitian case, the coherence will be destroyed once the $\mathcal{Q}$-symmetry is broken by coupling operator $\hat{\mathcal{O}}$ as shown in Fig.\ref{Fig3}.(b). This reflects the fact that coherence protected only by antiunitary symmetry is fragile under non-Hermitian perturbation. 

	\section{Summary and Outlook}
	\label{Summary}
	We study the fate of coherence in degenerate subspace which is protected by unitary symmetry or anti-unitary symmetry. With symmetries lying in system Hamiltonian, and the interaction between system and environment, we have demonstrated that the coherence could be fragile when the symmetry is anti-unitary. We elaborate on classification of various symmetry combinations and analyze the stability of coherence detailedly which is confirmed by several spin-3/2 models. The results of this investigation are extensions of Ref.\cite{Cooper} and could be applied to the investigation of robustness in time-reversal invariant topological edge states. 
	Recent experimental progresses in controlling and manipulating dissipation in ultracold atoms provide unprecedented opportunity for understanding the dynamics of open quantum system which is driven far from of equilibrium\cite{Bouganne,Tomita1,Tomita2,LuoL,DissHubbard,Takahashi1,Takahashi2}. We expect our work will guide the preparation for stable time-reversal symmetry protected topological states in laboratory.
	
	There is another intriguing issue about the open quantum systems beyond the Born-Markov approximation. In this case, the coherence protected by anti-unitary symmetry could also be fragile even the system couples hermitian operators with the same symmetry\cite{foot2}. 
	What's more, the current study focuses on non-interacting systems. An important issue in future study is exploring to the stability of quantum degeneracy protected by anti-unitary in interacting systems, especially to interacting topological phases in many-body system such as Haldane phase in AKLT model chain\cite{CaiZi}.
	

	\section*{Acknowledgements}
	We would like to thank Yu Chen for helpful discussions. This work is supported by Beijing Outstanding Young Scientist Program hold by Hui Zhai. 
	L. P acknowledges support from the project funded by the China Postdoctoral Science Foundation (Grant No. 2020M680496). 
	\appendix
	
	\section{Response of density matrix and von Neumann entropy growth}\label{App1}
	
	In this appendix, we derive the response of density matrix and von Neumann entropy growth by means of NHLRT.
	As discussed in Ref.\cite{NHLRT}, the whole system coupling to a bath with white noise can be described by a non-Hermitian effective Hamiltonian  
	\begin{eqnarray}
	\hat{H}_{\rm eff}=\hat{H}_S+\hat{H}_{\mathrm{diss}},\label{Ham_S1}
	\end{eqnarray}
	where $H_{\mathrm{diss}}=\left(-i\gamma\hat{\cal O}^\dag \hat{\cal O}^{}+\hat{\cal O}^\dag\hat{\xi}^{}+\hat{\xi}^\dag\hat{\cal O}^{}\right)$ and $\gamma=\pi|g|^2\rho$ is dissipation strength where $\rho$ is spectrum density of bath. 
	$\hat{\xi}(t)$,~$\hat{\xi}^\dag(t)$ present the Langevin noise operators which obey the following relations
	\begin{align}
	\label{xixidag}
	&\langle \hat{\xi}(t)\hat{\xi}^\dag(t_1)\rangle_{\rm noise}=2\gamma\delta(t-t_1),\nonumber\\
	&\langle \hat{\xi}(t)\hat{\xi}(t_1)\rangle_{\rm noise}=\langle \hat{\xi}^\dag(t)\hat{\xi}(t_1)\rangle_{\rm noise}=\langle \hat{\xi}^\dag(t)\hat{\xi}^\dag(t_1)\rangle_{\rm noise}=0,
	\end{align}
	where $\langle\cdots\rangle_{\rm noise}$ denotes the noise average \cite{foot1}. This formalism is equivalent to the total Hamiltonian with a white noise bath.  
	In the interaction picture, the time evolution of density matrix can be expressed by 
	
	\begin{eqnarray}
	\hat{\rho}(t)
	&=&\hat{U}_{\rm eff}(t)\hat{\rho}_0(t)\hat{U}^\dag_{\rm eff}(t)\nonumber \\ 
	\end{eqnarray}
	where
	$\hat{U}_{\rm eff}(t)={\rm \widetilde{T}}\exp{\left(-i\int_0^t \hat{H}_{\rm diss}(t')dt'\right)}$ with anti-time-ordered operator ${\rm \widetilde{T}}$ and $\hat{\rho}_0(t)=e^{-i \hat{H}_S t}\hat{\rho}(0)e^{i \hat{H}_S t}$ denotes the time-evolution of density matrix determined by $\hat{H}_S$. Taking $\hat{H}_{\mathrm{diss}}$ as perturbation and then averaging the noise, one can obtain the density matrix with the first-order correction of $\gamma$ 
	\begin{widetext}
		\begin{eqnarray}
		\rho(t)&\equiv&\left\langle \hat{\rho}(t)\right\rangle_{\rm noise}=\left\langle U'_{\rm eff}(t)\hat{\rho}_0(t)U'^\dag_{\rm eff}(t)\right\rangle_{\rm noise} \nonumber \\
		&=&\Bigg< \left(1+\sum_{n=1}^\infty(-i)^n\int_{t_1< \cdots< t_n}\hat{H}_{\rm diss}(t_1)\cdots\hat{H}_{\rm diss}(t_n)dt_1\cdots dt_n\right)\hat{\rho}_0(t)\nonumber \\
		&\times&\left(1+\sum_{n=1}^\infty(i)^n\int_{t_1< \cdots< t_n}\hat{H}^\dag_{\rm diss}(t_n)\cdots\hat{H}^\dag_{\rm diss}(t_1)dt_1\cdots dt_n\right)\Bigg>_{\rm noise}  \nonumber \\
		&\approx&\rho_0(t)-\int_{0}^{t}dt'\gamma\Big\{ \hat{\cal O}^\dag(t')\hat{\cal O}(t'),\rho_0(t)\Big\}+2\int_{0}^{t}dt'\gamma\hat{\cal{O}}(t')\rho_0(t)\hat{\cal O}^\dag(t'),
		\end{eqnarray}
	\end{widetext}
	where $\langle\cdots\rangle_{\rm noise}$ is the noise average and 
	the correlation function Eq.(\ref{xixidag}) has been applied. The linear response of density matrix is defined by 
	\begin{eqnarray}
	\delta\rho(t)\equiv\rho(t)-\rho_0(t)&=&-\int_{0}^{t}dt'\gamma\Big\{ \hat{\cal O}^\dag(t')\hat{\cal O}(t'),\rho_0(t)\Big\}\nonumber \\
	&+&2\int_{0}^{t}dt'\gamma\hat{\cal{O}}(t')\rho_0(t)\hat{\cal O}^\dag(t'). \label{delta_rho}
	\end{eqnarray}
	From the response of density matrix, one can calculate the von Neumann entropy which characterizes the loss of coherence.
	Here we focus on the density matrix in Krammers degenerate subspace $\rho_{\rm K}(t)=\Pi_{\rm K}\rho(t)\Pi_{\rm K}$ ($\rho_{0,\rm K}(t)=\Pi_{\rm K}\rho_0(t)\Pi_{\rm K}$) and the corresponding response of von Neumann entropy is given by 
	\begin{widetext}
		\begin{eqnarray}
		\delta S_{\rm v}(t)&=&S_{\rm v}(t)-S_{0,\rm R}(t)\nonumber \\
		&=&-\operatorname{Tr}\left[\frac{\rho_{\rm 0,K}(t)+\delta\rho_{\rm K}(t)}{\operatorname{Tr}\left(\rho_{\rm 0,K}(t)+\delta\rho_{\rm K}(t)\right)} \log \left(\frac{\rho_{\rm 0,K}(t)+\delta\rho_{\rm K}(t)}{\operatorname{Tr}\left(\rho_{\rm 0,K}(t)+\delta\rho_{\rm K}(t)\right)}\right)\right]+ \operatorname{Tr}\left[\rho_{\rm 0,K}(t) \log \rho_{\rm 0,K}(t)\right]\nonumber \\
		&=&\operatorname{Tr}\Big[\rho_{\rm 0,K}(t) \log \rho_{\rm 0,K}(t)-\left(\delta \rho_{\rm 0,K}(t) / \operatorname{Tr} \delta \rho_{\rm 0,K}(t)\right) \log \rho_{\rm 0,K}(t)\Big] \operatorname{Tr} \delta \rho_{\rm 0,K}(t).
		\end{eqnarray}
	\end{widetext}
	We initialize the density matrix in the Kramers' degenerate space, i.e. ${\rm Tr}\rho_{0,K}(0)=1$ which means ${\rm Tr}\rho_{0,K}(t)=1$.
	It is clearly that, when $\delta \rho_{\rm K}(t) \propto \rho_{\rm 0,K}(t)$ (namely, $\delta \rho_{\rm K}(t)=\rho_{\rm 0,K}(t) \operatorname{tr} \delta \rho_{\rm K}(t))$ we have $\delta S_{\rm v}(t)=0$, $\delta S_{\rm R}(t)=0$ but $\delta S_{\rm v}(t) \neq 0$ if $\delta \rho_{\rm K}(t) \neq \rho_{\rm 0,K}(t)$. In other words, the time-reversal symmetry breaking discussed in main text leads to the growth of entropy.
	
	\section{Matrix representation of quaternion group}\label{App2}
	This appendix provides matrix representation of quaternion group $\mathcal{Q}$ discussed in maintext. The $\mathcal{Q}$-group $\{Q_j,j=1,\cdots,8\}$ is a  non-Abelian group which is isomorphic to subset $\{1,i,j,k,-1,-i,-j,-k\}$ whose multiplication table is displayed in Table \ref{table2}.
	\begin{table}[!h]
		\centering
		\caption{The multiplication table (Cayley table) of quaternion group.}
		\label{table2}
		\begin{tabular}{ c c c c c c c c c }
			\hline
			\hline Element & $\mathrm{e}$ & $\overline{\mathrm{e}}$ & $i$ & $\overline{\mathrm{i}}$ & $j$ & $\overline{\mathrm{j}}$ & $k$ & $\overline{\mathrm{k}}$ \\
			\hline $\mathrm{e}$ & $\mathrm{e}$ & $\overline{\mathrm{e}}$ & $i$ & $\overline{\mathrm{i}}$ & $j$ & $\overline{\mathrm{j}}$ & $k$ & $\overline{\mathrm{k}}$ \\
			\hline-1 & -1 & $\mathrm{e}$ & $\overline{\mathrm{i}}$ & $i$ & $\overline{\mathrm{j}}$ & $j$ & $\overline{\mathrm{k}}$ & $k$ \\
			\hline $i$ & $i$ & $\overline{\mathrm{i}}$ & $\overline{\mathrm{e}}$ & $\mathrm{e}$ & $k$ & $\overline{\mathrm{k}}$ & $\overline{\mathrm{j}}$ & $j$ \\
			\hline$\overline{\mathrm{i}}$ & $\overline{\mathrm{i}}$ & $i$ & $\mathrm{e}$ & $\overline{\mathrm{e}}$ & $\overline{\mathrm{k}}$ & $k$ & $j$ & $\overline{\mathrm{j}}$ \\
			\hline$j$ & $j$ & $\overline{\mathrm{j}}$ & $\overline{\mathrm{k}}$ & $k$ & $\overline{\mathrm{e}}$ & $\mathrm{e}$ & $i$ & $\overline{\mathrm{i}}$ \\
			\hline$\overline{\mathrm{j}}$ & $\overline{\mathrm{j}}$ & $j$ & $k$ & $\overline{\mathrm{k}}$ & $\mathrm{e}$ & $\overline{\mathrm{e}}$ & $\overline{\mathrm{i}}$ & $i$ \\
			\hline$k$ & $k$ & $\overline{\mathrm{k}}$ & $j$ & $\overline{\mathrm{j}}$ & $\overline{\mathrm{i}}$ & $i$ & $\overline{\mathrm{e}}$ & $\mathrm{e}$ \\
			\hline$\overline{\mathrm{k}}$ & $\overline{\mathrm{k}}$ & $k$ & $\overline{\mathrm{j}}$ & $j$ & $i$ & $\overline{\mathrm{i}}$ & $\mathrm{e}$ & $\overline{\mathrm{e}}$ \\
			\hline
			\hline
		\end{tabular}
	\end{table}
	The $\mathcal{Q}$-group contains a two-dimensional irreducible representation (see Table.\ref{table3}) which can be described as a subgroup of the special linear group $\mathrm{SL}_{2}(\mathbb{C})$. We can construct the following 4-dimensional reducible representation 
	
	\begin{align}
	Q_1 &=I\otimes I,~Q_2 = -Q_1,~Q_3=-I\otimes i\sigma_z, ~Q_4=-Q_3,\nonumber \\
	Q_5&=-i\sigma_x\otimes\sigma_y,~Q_6=-Q_5,~Q_7=-i\sigma_x\otimes\sigma_x,~Q_8=-Q_7, \label{rep}
	\end{align}
	where $\sigma_{x,y,z}$ denote Pauli matrices and $I$ is identity matrix. One can easily find that the matrix representation (\ref{rep}) obeys the multiplication table (\ref{table2}) and contains two-dimensional irreducible representation.
	
	\begin{table}[!h]
		\centering
		\caption{The character table of quaternion group}
		\label{table3}
		\begin{tabular}{c c c c c c}
			\hline\hline Representation/Conjugacy class & $\{\mathrm{e}\}$ & $\{\overline{\mathrm{e}}\}$ & $\{\mathrm{i}, \overline{\mathrm{i}}\}$ & $\{\mathrm{j}, \overline{\mathrm{j}}\}$ & $\{\mathrm{k}, \overline{\mathrm{k}}\}$ \\
			Trivial representation & 1 & 1 & 1 & 1 & 1 \\
			i-kernel & 1 & 1 & 1 & -1 & -1 \\
			j-kernel & 1 & 1 & -1 & 1 & -1 \\
			k-kernel & 1 & 1 & -1 & -1 & 1 \\
			2-dimensional representation & 2 & -2 & 0 & 0 & 0 \\
			\hline\hline
		\end{tabular}
	\end{table}


\newpage
	\begin{thebibliography}{99}
		
		\bibitem{TopoClass1}A. P. Schnyder, S. Ryu, A. Furusaki, and A. W. W. Ludwig, Phys. Rev. B {\bf 78}, 195125 (2008).
		\bibitem{TopoClass2} A. Kitaev, AIP Conf. Proc. {\bf 1134}, 22 (2009).
		\bibitem{TopoClass3} S. Ryu, A. P. Schnyder, A. Furusaki, and A. W. W. Ludwig, New J. Phys. {\bf 12}, 065010 (2010).
		\bibitem{SymClass1}M. R. Zirnbauer, J. Math. Phys. (N.Y.) {\bf 37}, 4986 (1996).
		\bibitem{SymClass2}  A. Altland and M. R. Zirnbauer, Phys. Rev. B {\bf 55}, 1142 (1997).
		\bibitem{Topo1} M. Z. Hasan and C. L. Kane, Rev. Mod. Phys. {\bf 82}, 3045 (2010).
		\bibitem{Topo2} X.-L. Qi and S.-C. Zhang, Rev. Mod. Phys. {\bf 83}, 1057 (2011).
		\bibitem{Topo3} B. Bernevig and T. Hughes, Topological Insulators and Topological Superconductors (Princeton University Press,2013).
		\bibitem{Topo4} C.-K. Chiu, J. C. Y. Teo, A. P. Schnyder, and S. Ryu, Rev. Mod. Phys. {\bf 88}, 035005 (2016).
		\bibitem{NHTopo1} Z. Gong, Y. Ashida, K. Kawabata, K. Takasan, S. Higashikawa, and M. Ueda, Phys. Rev. X {\bf 8}, 031079 (2018).
		\bibitem{NHTopo2} K. Kawabata, K. Shiozaki, M. Ueda, and M. Sato, Phys. Rev. X {\bf 9}, 041015 (2019).
		\bibitem{NHTopo3} C.-H. Liu and S. Chen, Phys. Rev. B {\bf 100}, 144106 (2019).
		\bibitem{NHTopo4} H. Zhou and J. Y. Lee, Phys. Rev. B {\bf 99}, 235112 (2019).
		\bibitem{NHTopo5} K. Kawabata, T. Bessho, and M. Sato, Phys. Rev. Lett. {\bf 123}, 066405(2019).
		\bibitem{NHTopo6} K. Kawabata, S. Higashikawa, Z. Gong, Y. Ashida and M. Ueda, Nat Commun {\bf 10}, 297 (2019).
		\bibitem{NHTopo7} C.-H. Liu, H. Jiang, and S. Chen, Phys. Rev. B {\bf 99}, 125103 (2019).
		\bibitem{NHTopo8} J. Y. Lee, J. Ahn, H. Zhou, and A. Vishwanath,Phys. Rev. Lett. {\bf 123}, 206404 (2019).
		
		
		\bibitem{Cooper2} S. Lieu, M. McGinley, and N. R. Cooper, Phys. Rev. Lett. {\bf 124}, 040401 (2020). 
		\bibitem{Altland}A. Altland, M. Fleischhauer, S. Diehl, arXiv:2007.10448 (2020).
		\bibitem{Nielsen} M. A. Nielsen and I. L. Chuang, \textit{Quantum Computation and Quantum Information} (Cambridge University Press, 2010). 
		\bibitem{Wigner}E. Wigner, \textit{Gruppentheorie}  (Vieweg, Braunschweig, 1931).
		\bibitem{Cooper} M. McGinley and N. R. Cooper, Nat. Phys. {\bf 16}, 1181 (2020).
		
		
		
		\bibitem{NHLRT} L. Pan, X. Chen, Yu, Chen and H. Zhai, Nat. Phys. {\bf 16}, 767 (2020).
		
		\bibitem{NHLRT2} Tian-Shu Deng, Lei Pan, Yu Chen, Hui Zhai, arXiv:2009.13043 (2020).
		
		\bibitem{anti-Schur}J. O. Dimmock, J. Math. Phys. {\bf 4}, 1307 (1963).
		\bibitem{IrePre} Y. Jian and Z.-X. Liu, J. Phys. A {\bf 51}, 025207 (2017).
		\bibitem{foot1}It is worth emphasizing that the Langevin noise is essential since it preserves canonical commutation relation. Hamiltonian without Langevin noise contradicts the principle of quantum mechanics.  
		\bibitem{Choi}M.-D. Choi, Lin. Alg. Appl. {\bf 10}, 285 (1975).
		\bibitem{Jamiolkowski}A. Jamio{\l}kowski Rep. Math. Phys. {\bf 3}, 275 (1972).
		\bibitem{Schur}E. P. Wigner, \textit{Group Theory and Its Application to the Quantum Mechanics of Atomic Spectra} (Academic Press, New York, 1959).


		\bibitem{foot2} For the bath beyond Born-Markov approximation at finite temperature, the anti-unitary protected coherence in system could also be unstable against to hermitian coupling for the multi-channel case. This can be illustrated by a two-channel model with two independent coupling operators which give rise to a non-hermitian operator $c_1\mathcal{\hat{O}}_1+c_2\mathcal{\hat{O}}_2$ (with real numbers $c_1,c_2$) acting on the system. As discussed in section \ref{Anti-Uni-sym}, this non-Hermiticity leads to decoherence despite the possession of anti-symmetry. 
		\bibitem{NVcenter1} J. Choi, S. Choi, G. Kucsko, P. C. Maurer, B. J. Shields, H. Sumiya, S. Onoda, J. Isoya, E. Demler, F. Jelezko, N. Y. Yao, and M. D. Lukin, Phys. Rev. Lett. {\bf 118}, 093601 (2017).
		
		\bibitem{NVcenter2} G. Kucsko, S. Choi, J. Choi, P. C. Maurer, H. Zhou, R. Landig, H. Sumiya, S. Onoda, J. Isoya, F. Jelezko, E. Demler, N. Y. Yao, and M. D. Lukin,  Phys. Rev. Lett. {\bf 121}, 023601 (2018).

		\bibitem{Tomita1} T. Tomita, S. Nakajima, I. Danshita, Y. Takasu, and Y. Takahashi, Sci. Adv. {\bf 3}, e1701513 (2017).
		\bibitem{Tomita2} T. Tomita, S. Nakajima, Y. Takasu, and Y. Takahashi, Phys.Rev. A {\bf 99}, 031601(R) (2019).
		\bibitem{LuoL} J. Li, A. K. Harter, J. Liu, L. de Melo, Y. N. Joglekar, and L. Luo, Nat. Commun.{\bf 10}, 855 (2019).
		
		
		\bibitem{DissHubbard}K. Sponselee, L. Freystatzky, B. Abeln, M. Diem, B. Hundt, A. Kochanke, T. Ponath, B. Santra, L. Mathey, K. Sengstock, and C. Becker, Quantum Science and Technology {\bf 4}, 014002 (2019).
		
		\bibitem{Bouganne} R. Bouganne, M. B. Aguilera, A. Ghermaoui, J. Beugnon, and F. Gerbier,  Nat. Phys. {\bf 16}, 21 (2020). 
		\bibitem{Takahashi2} F. Sch\"{a}fer, T. Fukuhara, S. Sugawa, Y. Takasu, and Y. Takahashi, Nat. Rev. Phys. {\bf 2}, 411 (2020).
		\bibitem{Takahashi1} Yosuke Takasu, Tomoya Yagami, Yuto Ashida, Ryusuke Hamazaki, Yoshihito Kuno, Yoshiro Takahashi, arXiv:2004. 05734 (2020).
		\bibitem{CaiZi} Z. Wang, Q. Li, W. Li, Z. Cai, arXiv:2102.06383 (2021).
		
		
	\end{thebibliography}
\end{document}